\documentclass[12pt]{article} 
\usepackage[numbers]{natbib}
\usepackage{a4}
\usepackage{amsmath}
\usepackage{amssymb}
\usepackage{latexsym}
\def \pd{\partial}

\def \e{{\mathrm{e}}}

\def \BR{\boldsymbol{R}}

\def \Bs{{\boldsymbol{s}}}
\def \Bn{{\boldsymbol{n}}}

\def \rr{{\boldsymbol{r}}}

\begin{document}
%%%%%%%%%%%%%%%%%%%%%%%%%%%%%%%%%%%%%%%%%%%%%%%%%%%%%%%%%%%%%%%%%%%%%
\title{{\bf The solid angle and the Burgers formula 
in the theory of gradient elasticity: \\
line integral representation  
}}
\author{
Markus Lazar~$^\text{a,}$\footnote{
%%%%Corresponding author.
{\it E-mail address:} lazar@fkp.tu-darmstadt.de (M.~Lazar).
%%%\newline
%%%Tel.:~+49(0)6151/163686;~Fax.:~+49(0)6151/163681.
}
\
and Giacomo Po~$^\text{b}$
\\ \\
${}^\text{a}$ 
        Heisenberg Research Group,\\
        Department of Physics,\\
        Darmstadt University of Technology,\\
        Hochschulstr. 6,\\      
        D-64289 Darmstadt, Germany\\
\\
${}^\text{b}$
Mechanical and Aerospace Engineering,\\
University of California, Los Angeles,\\
Los Angeles, CA, 90095, USA
}

\date{\today}    
\maketitle
%%%%%%%%%%%%%%%%%%%%%%%%%%%%%%%%%%%%%%%%%%%%%%%%%%%%%%%%%%%%%%%%%%%%%%%%%%%%%

%%%\newpage

\begin{abstract}
A representation of the solid angle and the Burgers formula
as line integral is derived in the framework of 
the theory of gradient elasticity of Helmholtz type. 
The gradient version of the Eshelby-deWit representation 
of the Burgers formula of a closed dislocation loop is given.
Such a form is suitable for the numerical implementation 
in 3D dislocation dynamics (DD).\\

\end{abstract}
%%\vspace{1cm}

\section{Introduction}
The Burgers formula and the solid angle play an important role in
the dislocation theory~(e.g. \citep{deWit60,Lardner,HL,Teodosiu,LL}) 
and in the simulation of dislocation dynamics~(e.g. \citep{Ghoniem99,Ghoniem02,Li}).
The original formulas are given in the form of surface integrals.
The transformation of the surface integrals into line integrals 
was proposed by~\citet{deWit} and \citet{Eshelby} 
adopting Dirac's theory of magnetic monopoles~\citep{Dirac,Wentzel,BS}.
In particular, it turned out that the representation as line integral
is more appropriate for numerical implementation of these equations into the
dislocation dynamics. 
The classical expressions for the Burgers formula and for 
the solid angle are singular at the line of the dislocation loop. 
Moreover, the Burgers formula is discontinuous on the slip surface.

Non-singular expressions for the Burgers formula and the solid angle
have been recently found by~\citet{Lazar12,Lazar13} using
the theory of gradient elasticity of Helmholtz type.
The theory of gradient elasticity of Helmholtz type is a 
special  version 
of Mindlin's gradient elasticity theory
\citep{Mindlin64} (see also~\citep{LM05,Lazar13}) 
with only one characteristic length parameter.
\citet{LM06} have shown that, for straight dislocations, 
the gradient parameter leads to a smoothing of the displacement profile, 
in contrast to the jump occurring in the classical solution.
\citet{Lazar12,Lazar13} has given the generalized solid angle and
the corresponding part of the Burgers formula in the form of surface integrals.
In this letter, we recast the Burgers formula and the solid angle of 
gradient elasticity in compact form as line integrals over 
the closed dislocation loop. 
The results have a direct application to the numerical implementation 
and the computer simulation of non-singular dislocations 
within the so-called (discrete) dislocation dynamics.
In Section 2, we discuss and 
point out the basics of the line integral form of the solid angle
and of the associated vector potential
in the framework of classical elasticity and their relation to Dirac's
solution of a magnetic monopole. In Section 3, we derive the 
corresponding expressions in the framework of gradient elasticity.

\section{Classical elasticity}
In the theory of classical elasticity, 
the solid angle is given as a surface integral
(see, e.g.,~\cite{deWit60})
\begin{align}
\label{Omega-0}
\Omega^0(\rr)=\int_S v^0 _i(\BR)\, d S'_i
=\int_V v^0_i(\BR)\, \delta_i(S')\, d V'
=v^0_i(\rr) * \delta_i(S)
\,,
\end{align}
where the vector field $v_i^0$ is
\begin{align}
\label{v0}
v_i^0=-\frac{1}{2}\, \Delta \pd_i R
=-\pd_i \frac{1}{R}
=\frac{R_i}{R^3}\,,
\end{align}
while the Dirac $\delta$-function on a surface $S$~\citep{deWit2,Kleinert} is 
defined as
\begin{align}
\label{Dirac-S}
\delta_i(S)\equiv
\int_S \delta(\BR)\, d S_i'\,.
\end{align}
The relative radius vector $\BR=\rr-\rr'$ connects a source point $\rr'$ 
on the loop to a field point $\rr$ 
and $R=|\BR|$ denotes the norm of $\BR$.
Here $S$ denotes an arbitrary smooth surface enclosed by the loop $L$,
$d S'_i$ is an oriented surface element,
$\Omega^0(\rr)$ is the solid angle under which the loop $L$ is seen from 
the point $\rr$, and  $*$  denotes the spatial convolution.
The vector field~(\ref{v0}) is analogous to the magnetic field of
a magnetic monopole fixed at the origin (e.g.,~\citep{Wentzel,BS}).
The divergence  of the vector field~(\ref{v0}) yields
\begin{align}
\label{div-v0}
\pd_i v_i^0=-\frac{1}{2}\, \Delta \Delta\, R=4\pi\, \delta(\BR)\,,
\end{align}
since
\begin{align}
\Delta\Delta\, R=-8\pi\, \delta(\BR)\,.
\end{align}
The solid angle $\Omega^0$ is a multi-valued quantity with the residue $4\pi$.
Thus, the solid angle $\Omega^0$ changes by $4\pi$ when the field point 
crosses the the surface $S$. In particular, this happens for a Burgers circuit 
that  encircles $L$.
In other words, $S$ represents the surface of discontinuity.
Notice that, in classical elasticity, the plastic distortion caused 
by a dislocation loop is concentrated at the surface $S$. 
From a physical viewpoint,  $S$ represents the area swept 
by the loop $L$ during its motion and may be called the slip surface.
Thus, the surface $S$ is what determines the history of the plastic distortion
of a dislocation loop (see, e.g,~\citep{deWit60,deWit2}).

We may use the Stokes theorem to arrive at a line integral
over $L$ for the solid angle. 
To do so, it is necessary to express $v^0_i$ as the curl of a ``vector
potential"  $A^0_k$. However, Eq.~(\ref{div-v0}) shows that the divergence of
the vector field $v_i^0$ is not identically zero, and therefore it becomes
impossible to write $v^0_i$ everywhere
as the  curl of a vector potential. Nevertheless, introducing a
a so-called fictitious vector field 
$v_i^{(f)0}$, which is sometimes called ``string of singularity'', 
(see, e.g.,~\citep{Wentzel,BS})
having the property 
\begin{align}
\label{div-v0f}
\pd_i v_i^{(f)0}=-\pd_i v_i^{0}\,,
\end{align}
a vector potential $A^0_k$ 
may be introduced for the divergenceless sum $v_i^0+v_i^{(f)0}$:
\begin{align}
\label{A0-d}
v_i^0+v_i^{(f)0}=\epsilon_{ijk}\pd_j A^0_k\,.
\end{align}
Subtraction of the fictitious vector field in Eq.~(\ref{A0-d}) leads to the
physical vector field $v_i^0$ given by 
\begin{align}
\label{A0-d2}
v_i^0=\epsilon_{ijk}\pd_j A^0_k-v_i^{(f)0}\,.
\end{align}
Taking the curl of Eq.~(\ref{A0-d}) and imposing the ``Coulomb gauge''
$\pd_k A_k^0=0$, we find an inhomogeneous Laplace equation for the 
vector potential
\begin{align}
\label{A0-P}
\Delta A^0_k =-\epsilon_{klm}\pd_l v_m^{(f)0}\,,
\end{align}
where the fictitious vector field $v_m^{(f)0}$ is the source term of
the vector potential.
Using the 3D Green function of the 
Laplace equation, $-1/(4\pi R)$, the solution of Eq.~(\ref{A0-P}) reads
\begin{align}
\label{A0-Sol}
A^0_k(\rr)=\frac{1}{4\pi}\, \epsilon_{klm}
\int_V \pd_l\, \frac{1}{R}\, v_m^{(f)0}(\rr')\,d V'
=-\frac{1}{4\pi}\, \epsilon_{klm}
\int_V v^0_l(\BR)\, v_m^{(f)0}(\rr')\,d V'\,.
\end{align}

The fictitious singular vector field $v_i^{(f)0}$ can be taken as~\citep{BS,Kleinert}
\begin{align}
\label{v0f-C}
v_i^{(f)0}(\rr)=\int_C v_{k,k}^0(\rr-\Bs) d s_i=4\pi\int_C \delta(\rr-\Bs)\, d s_i
\equiv 4\pi\, \delta_i(C)
\,,
\end{align}
where $C$ is a curve, called the ``Dirac string'', 
starting at $-\infty$ and ending at the origin
and $\delta_i(C)$ is the $\delta$-function along the Dirac string.
The divergence of this field is concentrated at the endpoint of the string:
\begin{align}
\pd_i v_i^{(f)0}(\rr)=-4\pi\, \delta(\rr)=-\pd_i v_i^{0}(\rr)\,.
\end{align}
Then the vector potential of the monopole~(\ref{A0-Sol}) 
is given as a line integral along
the path $C$ (see, e.g.,~\citep{Wentzel}):
\begin{align}
\label{A0-C}
A_k^0(\rr)=\epsilon_{klm}\int_C v^0_m(\rr-\Bs)\, d s_l
=-\epsilon_{klm}\int_C \pd_m\, \frac{1}{|\rr-\Bs|}\, d s_l\,.
\end{align}
The fictitious vector field $v_m^{(f)0}$ is a singular field which vanishes 
everywhere except along the Dirac string $C$.

If we choose for the path $C$ a straight line in the direction 
of a constant unit vector $n_i$, the fictitious vector field reads
\begin{align}
\label{v0f}
v_i^{(f)0}(\rr)=4\pi\,n_i \int_{-\infty}^0 \delta(\rr-\Bn s)\, d s\,,
\end{align}
and the vector potential 
of the ``magnetic monopole'' reduces to 
\begin{align}
\label{A0}
A_k^0(\rr)=\epsilon_{klm}\,\frac{n_l r_m}{r(r+r_i n_i)}\,.
\end{align}
%%where $n_i$ is an arbitrary but constant unit vector
%%which shows the lack of uniqueness.
Eq.~(\ref{A0}) has the original form of the vector potential of
Dirac's magnetic monopole (see, e.g.,~\cite{Dirac,Wentzel})
which was adopted by~\citet{deWit} and \citet{Eshelby}
for the representation of the solid angle as a line integral.

Substituting Eqs.~(\ref{A0-d2}) and (\ref{v0f-C}) into (\ref{Omega-0}) and using
the Stokes theorem, we find
\begin{align}
\label{Omega-0-1}
\Omega^0(\rr)
&=v^0_i(\rr) * \delta_i(S)
=\epsilon_{ijk}\pd_j A^0_k(\rr)*\delta_i(S)-v_i^{(f)0}(\rr)*\delta_i(S)
\nonumber\\
&=A^0_k(\rr)*\epsilon_{kji}\pd_j\delta_i(S)-4\pi\,\delta_i(C)*\delta_i(S)
\nonumber\\
&=A^0_k(\rr)*\delta_k(L)-4\pi\,\delta_i(C)*\delta_i(S)
\,,
\end{align}
where 
\begin{align}
A^0_k(\rr)*\delta_k(L)
=\int_V A^0_k(\BR) \delta_k(L')\, d V'
=\oint_L A^0_k(\BR)\, d L_k'\,,
\end{align}
the $\delta$-function on a closed line $L$~\citep{deWit2,Kleinert}
\begin{align}
\label{Dirac-L}
\delta_i(L)\equiv
\int_L \delta(\BR)\, d L_i'\,,
\end{align}
and $\epsilon_{kji}\pd_j\delta_i(S)=\delta_k(L)$.
Here $d L'_i$ denotes the line element at $\rr'$.
For the contribution of the fictitious vector field 
we used the formula~\citep{deWit2,Teodosiu70}
\begin{align}
%%%1_{LS}\equiv
\delta_i(L)*\delta_i(S)=
\int_S\int_L\delta(\rr-\rr')\, d L'_i\, d S_i
=
\left\{ 
\begin{array}{rl}
\displaystyle{1}\,,\qquad &\text{if $L$ crosses $S$ positively}\\
\displaystyle{0}\,,\qquad &\text{if $L$ does not cross $S$}\\
\displaystyle{-1}\,,\qquad &\text{if $L$ crosses $S$ negatively}\\
\end{array}
\right.
\ .
\end{align}
Finally, the solid angle reduces to a line integral 
of the monopole vector potential~(\ref{A0-C}) or (\ref{A0}) and a constant
\begin{align}
\label{Omega-0-2}
\Omega^0(\rr)=\oint_L A^0_k(\BR)\, d L'_k
-4\pi
%%%\, 1_{CS}
\left\{ 
\begin{array}{rl}
\displaystyle{1}\,,\qquad &\text{if $C$ crosses $S$ positively}\\
\displaystyle{0}\,,\qquad &\text{if $C$ does not cross $S$}\\
\displaystyle{-1}\,,\qquad &\text{if $C$ crosses $S$ negatively}\\
\end{array}
\right.
\ .
\end{align}
We mention that~\citet{deWit} and \citet{Eshelby} have neglected the
subtraction of 
the fictitious vector field of the semi-infinite solenoid of the monopole field
in their calculations. Of course, if 
it is  chosen that the path $C$ of the solenoid
does not cross $S$, the fictitious vector field does not give a contribution to the solid angle in Eq.~(\ref{Omega-0-2}).
An equation like~(\ref{Omega-0-2}) was also derived by~\citet{Asvestas}
for physical optics.

\section{Gradient elasticity}
In the theory of gradient elasticity of Helmholtz type, 
the solid angle is defined as the following
surface integral~\citep{Lazar12,Lazar13}
\begin{align}
\label{Omega}
\Omega(\rr)=\int_S v_i(\BR)\, d S'_i
=\int_V v_i(\BR)\, \delta_i(S')\, d V'
=v_i(\rr) * \delta_i(S)\,,
\end{align}
where the non-singular vector field is given by\footnote{
It interesting to note,
from the historical point of view, that also \citet{PP} 
considered such a non-singular vector field~(\ref{v}) for a solid angle.}
\begin{align}
\label{v}
v_i=-\frac{1}{2}\, \Delta \pd_i\, A(R)
=-\pd_i \bigg[\frac{1}{R}\big(1-\e^{-R/\ell}\big)\bigg]
=\frac{R_i}{R^3}\Big(1-\Big(1+\frac{R}{\ell}\Big)\e^{-R/\ell}\Big)
\end{align}
with the ``regularization function''
\begin{align}
\label{A-R}
A(R)=R+\frac{2\ell^2}{R}\big(1-\e^{-R/\ell}\big)\,,
\end{align}
and 
$\ell$ denotes the characteristic internal length of gradient elasticity.
In the framework of gradient elasticity, 
the displacement vector of a closed dislocation loop in an isotropic
elastic 
material is given by the generalized Burgers formula~\citep{Lazar12,Lazar13}
\begin{align}
\label{u-Burger-grad-S}
u_i(\rr) = -\frac{b_i}{4\pi}\,\Omega(\rr)
+\frac{b_l\epsilon_{klj}}{8\pi} \oint_L
\bigg\{\delta_{ij} \Delta -\frac{1}{1-\nu}\, \pd_i \pd_j \bigg\}\, 
A(R)\,  d L'_k\, ,
\end{align}                  
where $\mu$ and $\nu$ are the shear modulus and Poisson's ratio, respectively,
and $b_i$ is the Burgers vector.

The function $A(R)$ satisfies the following relations~\cite{Lazar13}
\begin{align}
\label{A-LDD}
L\,\Delta\Delta\, A(R)&=-8\pi\, \delta(\BR)\,,\\
\label{A-DD}
\Delta\Delta\, A(R)&=-8 \pi\, G(R)\,,\\
\label{A-L}
L\, A(R)&=R\,,
\end{align}
with the Helmholtz operator
\begin{align}
\label{L}
L=1-\ell^2\Delta\,.
\end{align}
The Green function of the three-dimensional Helmholtz equation
is defined by
\begin{align}
L\, G(R)&=\delta(\BR)\,,\qquad
\label{G}
G(R)=\frac{\e^{-R/\ell}}{4\pi \ell^2 R}\,.
\end{align}
%%%and $\int G(R)\,d V'=1$.

If we multiply Eq.~(\ref{v}) by the Helmholtz operator~(\ref{L}) and 
use the relation~(\ref{A-L}), 
we obtain
\begin{align}
\label{Lv}
L\, v_i=-\frac{1}{2}\, \Delta \pd_i\,L\, A(R)=-\frac{1}{2}\, \Delta \pd_i\,R\,.
\end{align}
Comparing Eqs.~(\ref{v0}) and (\ref{Lv}), we find 
an inhomogeneous Helmholtz equation for $v_i$:
\begin{align}
\label{v-H}
L\, v_i=v_i^0\,,
\end{align}
where $v_i^0$ gives the inhomogeneous part.
Using Eq.~(\ref{A-DD}),
the divergence of the vector field~(\ref{v}) is calculated as
\begin{align}
\label{div-v}
\pd_i v_i=-\frac{1}{2}\, \Delta \Delta\, A(R)=4\pi\, G(R)\,.
%% \qquad
\end{align}
Multiplying Eq.~(\ref{div-v}) by the Helmholtz operator $L$
and using Eqs.~(\ref{A-LDD}) and (\ref{G}) or performing the divergence
of Eq.~(\ref{v-H}), we get
\begin{align}
\label{Ldiv-v}
L\, \pd_i v_i=-\frac{1}{2}\, L \Delta \Delta\, A(R)
=4\pi\, L\,G(R)= 4\pi\, \delta(\BR)\,.
\end{align}

In order to introduce a ``vector potential'' $A_k$ in the framework 
of gradient elasticity,  we 
introduce a fictitious vector field $v_i^{(f)}$ with the property that
\begin{align}
\label{div-vf}
\pd_i v_i^{(f)}=-\pd_i v_i\,.
\end{align}
A vector potential can now be introduced for the divergenceless sum $v_i+v^{(f)}_i$:
\begin{align}
\label{A-d}
v_i+v^{(f)}_i=\epsilon_{ijk}\pd_j A_k\,.
\end{align}
The field $A_k$ in~(\ref{A-d}) is a monopole vector field in gradient theory.
If we subtract the fictitious vector field, we obtain the vector field 
\begin{align}
\label{A-d2}
v_i=\epsilon_{ijk}\pd_j A_k-v^{(f)}_i\,,
\end{align}
which is independent of the Dirac string. Only,
the vector potential $A_k$ and the fictitious vector field $v^{(f)}_i$
depend on the Dirac string $C$.
If we now take the curl of Eq.~(\ref{A-d}) and impose
the ``Coulomb gauge'' $\pd_k A_k=0$, we find 
that the vector potential satisfies the following inhomogeneous Laplace equation
\begin{align}
\label{A-P}
\Delta A_k =-\epsilon_{klm}\pd_l v_m^{(f)}\,,
\end{align}
and therefore has solution
\begin{align}
\label{A-Sol}
A_k(\rr)=\frac{1}{4\pi}\, \epsilon_{klm}
\int_V \pd_l\, \frac{1}{R}\, v_m^{(f)}(\rr')\,d V'\,.
\end{align}

Multiplying Eq.~(\ref{A-d}) by the Helmholtz operator $L$ and
using Eqs.~(\ref{v-H}) and (\ref{A0-d}), 
we obtain an inhomogeneous Helmholtz equation for
the vector potential $A_k$
\begin{align}
\label{LA}
L A_k=A_k^0\,,
\end{align}
where the right hand side is given by the singular monopole field $A_k^0$,
and an inhomogeneous Helmholtz equation for
the fictitious vector field $v_i^{(f)}$
\begin{align}
\label{Lvf}
L\, v^{(f)}_i=v_i^{(f)0}\,
\end{align}
with $v_i^{(f)0}$ as inhomogeneous part.
If we multiply Eq.~(\ref{A-P}) by the Helmholtz operator $L$ 
and use Eq.~(\ref{Lvf}),
we obtain an inhomogeneous Helmholtz-Laplace equation for the
vector potential
\begin{align}
\label{A-PL}
L\Delta\, A_k =-\epsilon_{klm} \pd_l v_m^{(f)0}\,,
\end{align}
where $v_i^{(f)0}$ is now the source of the vector potential 
of gradient elasticity.
The solution of~(\ref{A-PL}) is given by
\begin{align}
\label{A-SolL}
A_k(\rr)=\frac{1}{4\pi}\, \epsilon_{klm}
\int_V \pd_l\, 
\frac{1}{R}\Big(1-\e^{-R/\ell}\Big) v_m^{(f)0}(\rr')\,d V'
=-\frac{1}{4\pi}\, \epsilon_{klm}
\int_V v_l(\BR)\,  v_m^{(f)0}(\rr')\,d V'\,.
\end{align}

Moreover, the solution of Eq.~(\ref{LA}) might be written as 
a convolution integral
\begin{align}
\label{AC}
A_k=G*A_k^0\,,
\end{align}
which is a regularization of the singular monopole vector potential $A_k^0$.
Thus, it gives a non-singular monopole vector potential.
The solution of Eq.~(\ref{Lvf}) is 
\begin{align}
\label{vfC}
v^{(f)}_i=G*v_i^{(f)0}\,.
\end{align}

Using the Green function~(\ref{G}) of the 3D Helmholtz equation, 
the formal solutions~(\ref{AC}) and (\ref{vfC}) read explicitly
\begin{align}
\label{AC2}
A_k(\rr)=\frac{1}{4\pi\ell^2} 
\int_V \frac{\e^{-R/\ell}}{R}\, A^0_{k}(\rr')\,d V'\,,
\end{align}
and 
\begin{align}
\label{vf2}
v_i^{(f)}(\rr)=\frac{1}{4\pi\ell^2} 
\int_V \frac{\e^{-R/\ell}}{R}\, v^{(f)0}_{i}(\rr')\,d V'\,.
\end{align}
It is interesting to note that the solution~(\ref{AC2}) is similar in the
formal form to the solution of a ``massive'' magnetic monopole in 
massive electrodynamics given in~\citep{IJ}.

Substituting Eqs.~(\ref{v0f-C}) and (\ref{v0f}) into the
integral~(\ref{vf2}) and performing the integration in $V'$, 
the following forms of the fictitious vector field are obtained
\begin{align}
\label{vf-sol}
v^{(f)}_i(\rr)=\frac{1}{\ell^2} 
\int_C \frac{\e^{-|\rr-\Bs|/\ell}}{|\rr-\Bs|}\,d s_i
=4\pi\, G*\delta_i(C)
\end{align}
and
\begin{align}
\label{vf-sol2}
v^{(f)}_i(\rr)=\frac{n_i}{\ell^2} 
\int_{-\infty}^0 \frac{\e^{-|\rr-\Bn s|/\ell}}{|\rr-\Bn s|}\,d s\,,
\end{align}
respectively.
Solutions~(\ref{vf-sol}) and (\ref{vf-sol2}) consist of a kernel in 
the form of a thin tube along the Dirac string decreasing exponentially 
outside this kernel instead of the classical $\delta$-string.

The gradient solution of Eq.~(\ref{LA}) 
with the classical monopole fields~(\ref{A0-C}) and (\ref{A0})
is given by
\begin{align}
\label{A-C}
A_k(\rr)=\epsilon_{klm}\,\int_C v_m(\rr-\Bs)\, d s_l
=-\epsilon_{klm}\,\int_C \pd_m\, \frac{1}{|\rr-\Bs|}
\Big(1-\e^{-|\rr-\Bs|/\ell}\Big)\, d s_l
\end{align}
and
\begin{align}
\label{A-C2}
A_k(\rr)=\epsilon_{klm}\,n_l \int_{-\infty}^0 v_m(\rr-\Bn s)\, d s
=\epsilon_{klm}\, n_l \bigg(\frac{r_m}{r(r+r_i n_i)}
+ \pd_m  \int_{-\infty}^0  \frac{\e^{-|\rr-\Bn s|/\ell}}{|\rr-\Bn s|}
\, d s\bigg)\,.
\end{align}
Therefore, Eqs.~(\ref{A-C}) and (\ref{A-C2}) are the regularized 
versions of the singular vector potentials~(\ref{A0-C}) and (\ref{A0}). 
In general, Eqs.~(\ref{A-C}) and (\ref{A-C2}) are the non-singular
version of a Dirac monopole valid in gradient theory.
It is noted that Eqs.~(\ref{v0f-C}) and (\ref{v0f}) 
may be substituted directly 
into Eq.~(\ref{A-SolL}) in order to obtain Eqs.~(\ref{A-C})
and (\ref{A-C2}), respectively.

Substituting Eqs.~(\ref{A-d2}), (\ref{vfC}) and (\ref{v0f-C})
into (\ref{Omega}) and using
the Stokes theorem, we obtain
\begin{align}
\label{Omega-1}
\Omega(\rr)
&=v_i(\rr) * \delta_i(S)
=\epsilon_{ijk}\pd_j A_k(\rr)*\delta_i(S)-v_i^{(f)}(\rr)*\delta_i(S)
\nonumber\\
&=A_k(\rr)*\epsilon_{kji}\pd_j\delta_i(S)-G(\rr)*v_i^{(f)0}(\rr)*\delta_i(S)
\nonumber\\
&=A_k(\rr)*\delta_k(L)-4\pi\,G(\rr)*\delta_i(C)*\delta_i(S)
\,,
\end{align}
where 
\begin{align}
A_k(\rr)*\delta_k(L)
=\int_V A_k(\BR) \delta_k(L')\, d V'
=\oint_L A_k(\BR)\, d L_k'
\end{align}
and
\begin{align}
\label{Fict-int}
G(\rr)*\delta_i(C)*\delta_i(S)
=\int_S\int_C G(\rr-\rr')\, d L'_i\, d S_i
\,.
\end{align}
Multiplying Eq.~(\ref{Omega-1}) by the 
Helmholtz operator~(\ref{L}), 
it yields the relation between the solid angle in classical elasticity
and gradient elasticity
\begin{align}
\label{LOmega}
L\,\Omega(\rr)
&=L\,A_k(\rr)*\delta_k(L)-4\pi\,L\,G*\delta_i(C)*\delta_i(S)
\nonumber\\
&=A^0_k(\rr)*\delta_k(L)-4\pi\, \delta_i(C)*\delta_i(S)
=\Omega^0(\rr)\,.
\end{align}
Finally, the solid angle reduces to a line integral 
of the monopole vector potential~(\ref{A-C}) or (\ref{A-C2}) and 
a contribution due to the fictitious vector field $v_i^{(f)}$
\begin{align}
\label{Omega-2}
\Omega(\rr)=\oint_L A_k(\BR)\, d L'_k
-4\pi\int_S\int_C G(\rr-\rr')\, d L'_i\, d S_i\, ,
\end{align}
where $G$ is the Green function of the three-dimensional Helmholtz
equation~(\ref{G}).
$G(\rr-\rr')$ is non-zero for $\rr=\rr'$ and for $\rr$ different than $\rr'$
near the Dirac string. 
Therefore, the contribution of the fictitious vector  is
not localized at the intersection of the slip surface and 
the Dirac string, in contrast with the classical theory. 
 It is a continuous contribution which does not give rise to a jump of the slip surface.
 This result is consistent with the surface representation of the generalized solid angle~(\ref{Omega}).
 In gradient elasticity, the solid angle is non-singular and gives rise 
to a smoothing of the displacement profile in the Burgers formula.

It is noteworthy that if the  line direction of the Dirac string can be made
orthogonal to the slip surface, then the contribution of the fictitious vector
field vanishes identically. 
This condition is verified for common cases of practical interest. For
example, for dislocation loops gliding on planes the slip surface is flat and
the Dirac string can be chosen to lie on a plane parallel to the slip plane. 
For prismatic loops spanning a glide
cylinder,  the Dirac string can be chosen to be the cylinder axis.  Under
these and other conditions,  
the solid angle~(\ref{Omega}) 
may be transformed into a line integral of the ``vector potential''~(\ref{AC2})
\begin{align}
\label{Omega-3}
\Omega(\rr)=\oint_L A_k(\BR)\, d L'_k\,
\end{align}
and the corresponding Burgers formula~(\ref{u-Burger-grad-S}) 
can be expressed as a line integral performed over the closed dislocation loop
$L$ as
\begin{align}
\label{u-Burger-grad}
u_i(\rr) = 
-\frac{b_i}{4\pi}\oint_L A_k(\BR)\, d L'_k 
+\frac{b_l\epsilon_{klj}}{8\pi} \oint_L
\bigg\{\delta_{ij} \Delta -\frac{1}{1-\nu}\, \pd_i \pd_j \bigg\}\, 
A(R)\,  d L'_k\, .
\end{align}
Eq.~(\ref{u-Burger-grad}) is the line integral representation
of the Burgers formula in the framework of gradient elasticity of
Helmholtz type.
It enjoys three fundamental properties that make it particularly appealing for
numerical implementation in dislocation dynamics codes
(e.g.~\citep{Ghoniem99,Ghoniem02,Li}). First, it is non-singular on $L$,
second it is continuous on $S$, and third it involves only line integrals
along the dislocation loop. Applications of the theory presented in this
article  will be presented elsewhere \citep{Po13}.
Finally, Eq.~(\ref{u-Burger-grad}) may be called the gradient version 
of the Eshelby-deWit representation of the Burgers formula 
of a closed dislocation loop.

\section*{Acknowledgement}
M.L. gratefully acknowledges the grants from the 
Deutsche Forschungsgemeinschaft (Grant Nos. La1974/2-1, La1974/2-2, La1974/3-1).

\end{document}